\documentclass[twocolumn,showpacs,preprintnumbers,amsmath,amssymb]{revtex4}
\usepackage{graphicx}
\usepackage{dcolumn}
\usepackage{braket}

\renewcommand{\vec}{\boldsymbol}

\newcommand{\avr}[1]{\braket{#1}}

\newcommand{\kB}{\ensuremath{k_\text{B}}}

\newcommand{\relint}{\ensuremath{\lambda}}

\newcommand{\uL}{\ensuremath{u_\text{L}}}
\newcommand{\ur}{\ensuremath{u_\text{rel}}}
\newcommand{\urm}{\ensuremath{\bar u_\text{rel}}}

\newcommand{\Ecorr}{\ensuremath{C_E}}

\begin{document}
\bibliographystyle{apsrev}


\title{Melting of trapped few particle systems}

\author{J.~B\"oning$^1$}
\author{A.~Filinov$^1$}
\author{P.~Ludwig$^1$}
\author{H.~Baumgartner$^1$}
\author{M.~Bonitz$^1$}
\author{Yu.E.~Lozovik$^2$}
\affiliation{$^1$Institut f\"ur Theoretische Physik und Astrophysik, Christian-Albrechts-Universit\"{a}t zu Kiel, D-24098 Kiel, Germany}
\affiliation{$^2$Institute for Optics and Spectroscopy of the RAS, Troitsk, Russia}


\date{\today}
\begin{abstract}
In small confined systems predictions for the melting point strongly depend on the choice of quantity and on the way it is computed, even yielding divergent and ambiguous results. We present a very simple quantity which allows to control these problems -- the variance of the block averaged interparticle distance fluctuations. 
\end{abstract}

\maketitle
Crystallization and melting and, more generally, phase transitions are well known to pertain to very large systems only. At the same time, solid-like or liquid-like behavior has been observed in finite systems containing only one hundred or even $10$ particles and is becoming of increasing interest in many fields of physics, chemistry, and beyond. Current examples include bosonic crystals and supersolids, e.g{.} \cite{overview}, electrons or excitons in quantum dots \cite{filinov-etal.01prl}, ions in traps \cite{itano}, dusty plasma crystals \cite{bonitz-etal.prl06}, atomic clusters \cite{frantz,proykova06}, polymers \cite{berry02} etc. The notion of liquid and solid ``phases'' has been used successfully to characterize qualitatively different behaviors which resemble the corresponding properties in macroscopic systems and will be used here as well, following the definition of ref. \cite{proykova06}.
From the existence of phase-like states in very small systems arises the fundamental question of how to characterize phase changes and further, how many particles does it take at least to observe a phase transition. 

In macroscopic systems a solid-liquid transition can be verified by a variety of quantities including free energy differences, order parameters, specific heat, transport properties, structure factors, correlation functions and so on, e.g{.}~\cite{proykova06,binder,loewen94} which yield more or less equivalent results for the melting point. A particularly simple and transparent quantity is magnitude of the particle position fluctuations normalized to the interparticle distance (Lindemann ratio \uL{}), for an overview see. \cite{frenkel91,loewen94}.
But when applied to two-dimensional (2D) systems, \uL{} shows a logarithmic divergence with system size \cite{mermin}. This led to modified definitions, including the \emph{relative interparticle distance fluctuations} (IDF) \cite{bedanov85,etters75,berry88}
\begin{equation}
	\ur=\frac{2}{N (N-1)}\sum_{1\le i<j}^N \sqrt{ \frac{\avr{r_{ij}^2}}{\avr{r_{ij}}^2} - 1 },
	\label{ur}
\end{equation}
which are also well behaved in macroscopic 2D and 1D systems. Here $r_{ij}=|\vec r_i - \vec r_j|$ is the distance between two particles and $\avr{\ldots}$ denotes thermal averaging. In macrosopic systems \ur{} shows a jump at the melting point which clearly reflects the increased delocalization of particles in the liquid phase compared to a crystal.

However, when applied to small systems, $N < 100$, neither \uL{} nor \ur{} exhibit a jump upon classical or quantum melting, but rather a continuous increase over some finite temperature or density interval~\cite{filinov-etal.01prl} -- a familiar finite size effect. Therefore, it is very difficult to determine a transition point and the critical magnitude of the fluctuations $\ur^{\text{crit}}$. 
Even worse, the result for \ur{} (and hence the melting point) depend crucially on the method of calculation and on its duration. Increasing the length of a simulation (and the expected accuracy) may lead to growing systematic errors predicting a too low melting temperature, as was noted by Frantz \cite{frantz} and a few others. This is, of course, critical for reliable computer simulation of phase transitions in finite systems.
In this Letter, we analyze the reasons of this behavior and present a solution. We propose a novel quantity, the \emph{variance of the block averaged interparticle distance fluctuations}, which is sensitive to melting transitions and does not exhibit the convergence problems of \ur{}. We demonstrate the behavior of this quantity both, for classical and quantum melting by performing classical Monte Carlo (MC) and path integral Monte Carlo (PIMC) simulations, respectively. 

\begin{figure}
 \includegraphics[width=0.45\textwidth]{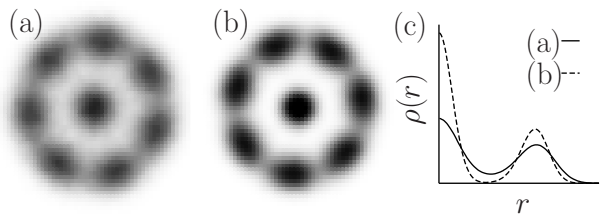}
 \caption{Configuration of a 2D trapped quantum system of $N=8$ spin-polarized bosons described by Eq.~\eqref{h}. Fig.~a) shows the liquid state ($\relint_1=14$) and b) the solid state ($\relint_4=30$), temperature is close to the ground state. Figure c) shows the radial density profile $\rho(r)$ for both configurations.}
 \label{fig:-1}
\end{figure}

{\bfseries Model and parameters.} While our approach is generally applicable we concentrate on strongly correlated classical or quantum particles in a parabolic trap in 2D and 3D described by the Hamiltonian
\begin{equation}
	\hat H = \sum_{i=1}^N \frac{\hat{\vec p}_i^2}{2m} + \sum_{i=1}^N \frac{m}{2}\omega^2 \vec r_i^2 +\sum_{1\le i<j}^N \frac{e^2}{|{\vec r}_i - {\vec r}_j|}.
	\label{h}
\end{equation}
The system is in a heat bath with temperature $T$ and has a fixed particle number $N$ (canonical ensemble). Below we use the dimensionless temperature $\kB Tr_0/e^2\to T$ where $r_0$ denotes the ground state distance of two particles, $r_0^3=2e^2/m\omega^2$. For quantum systems, the coupling parameter is $\relint=e^2/(l_0\hbar \omega)$ where $l_0$ is the oscillator length $l_0^2=\hbar/(m\omega)$. The ground state of this system consists of concentric spherical rings (2D), cf. Fig.~\ref{fig:-1} or shells (3D) \cite{itano,bonitz-etal.prl06}. This model has been very successul in describing trapped particles in many fields and has the advantage that classical melting (by temperature increase) and quantum melting (via compression by increasing $\omega$), including spin effects \cite{filinov-etal.01prl,afilinov_bose07}, can be analyzed on equal footing \cite{notesurface}.

{\bfseries Liquid and solid ``phases''.} 
The potential energy landscape of the system (\ref{h}) has numerous local minima but, in contrast to other finite systems such as atomic clusters \cite{berry02,proykova06}, they are not associated with ``phases'' but rather correspond to the ground state and metastable states (e.g{.} different shell configurations) which often are energetically very close, e.g.~\cite{ludwig05}. With increasing temperature an increasing number of these states becomes occupied. Melting proceeds as an isomerization transition with the system switching rapidly between a fast growing number of different configurations above some threshold temperature~\cite{baletto05}. 

\begin{figure}
 \includegraphics[width=0.45\textwidth]{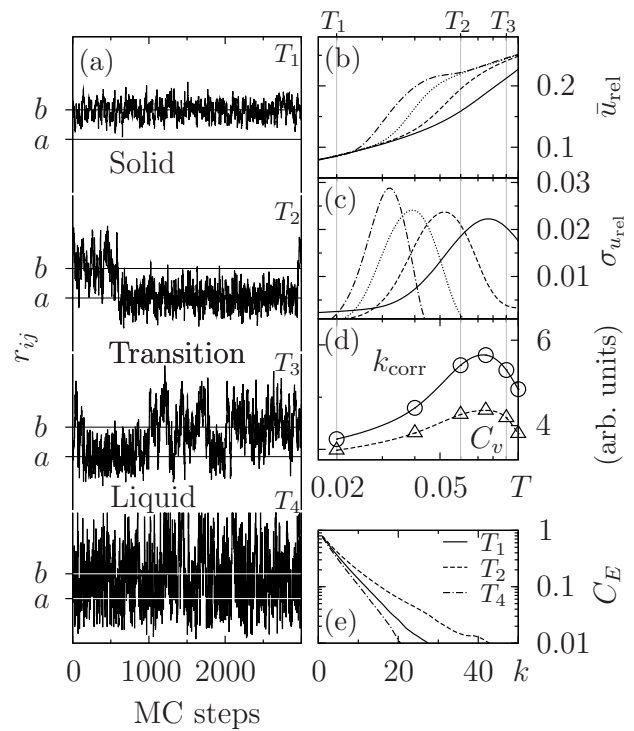}
 \caption{(a) Distance of an arbitrary pair of $N=4$ classical particles in 2D as a function of MC step. From top to bottom: $T_1=0.02$ (solid-like), $T_2=0.06$ and $T_3=0.09$ (transition region), and $T_4=0.5$ (liquid-like). $a$ and $b=\sqrt{2}a$ denote the two possible interparticle distances in the ground state. (b) Temperature dependence of the mean block averaged IDF \urm, for different block lengths $M=10^3,10^4,10^5,10^6$ (right to left) [equivalent to computing \ur{}, Eq.~\eqref{ur}, from multiple simulations of length $L=M$]. (c) The corresponding second moment $\sigma_{\ur}$, Eq.~\eqref{su}. (d) Specific heat $C_v$ and energy correlation time $k_{\text{corr}}$. (e) Total energy autocorrelation function \Ecorr{}, Eq.~\eqref{e}, for three temperatures.
}
 \label{fig:0}
\end{figure}
However, with reduction of $N$ the number of stationary states decreases until only the ground state remains. This is the case for $N=4$ in 2D which, due to its simplicity, allows for a transparent analysis of melting processes in the system \eqref{h}. The pair distances show a characteristic behavior as a function of simulation time (MC step) $k$, cf. Fig.~\ref{fig:0}.a): oscillations around some average value followed by a jump to a different value and again oscillations around a different mean and so on. This is readily understood: in its ground state the particles occupy the corners of a square of length $a$, so there exist two possible values for the six pair distances: $a$ and the diagonal $b=\sqrt{2}a$, which are the mean values around which the distances fluctuate. A \emph{jump} occurs whenever two particles $i$ and $j$ exchange their positions. Then the distances $r_{ik}$ and $r_{jk}$ to the remaining particles will change. While this leads to the same ground state (permutational isomer) this process costs energy associated with overcoming of a potential energy barrier. With increasing temperature, the frequency $\nu_j$ of these jumps grows steadily until around  $T=T_2$ a rapid growth of $\nu_j$ is observed. Finally, at $T=T_4$, pair exchanges occur constantly (bottom of Fig.~\ref{fig:0}.a), and particles are practically delocalized.
This behavior of $\nu_j$ clearly resembles a ``phase transition'' with the melting point being located inbetween the two limits $T_1$ (solid) and $T_4$ (liquid).

We verify this hypothesis by computing the IDF, Eq.~\eqref{ur} for this system, cf. Fig.~\ref{fig:0}.b). At low temperatures, \ur{}  is small, slowly increasing with $T$. Around $T=T_2$ the increase steepens slightly (rightmost curve). Repeating the calculations with higher accuracy, by subsequently increasing the simulation length $L$ (number of MC-steps) by factors $10, 100, 1000$, \ur{} shifts left towards smaller temperatures, and no convergence is observed. Thus, longer calculations yield an increase of \ur{} already in the solid-like regime, even though jumps are very rare, so the results for \ur{}, Eq.~\eqref{ur}, are ambiguous and unreliable. The reason is that, even in the solid state, a jump will be captured if $L$ is sufficiently long. This immediately leads to a significant increase of \ur{} emulating liquid-like behavior~\cite{noteurel}. Similar observations were made for clusters in Ref.~\cite{frantz}.

\begin{figure}
 \includegraphics[width=0.44\textwidth]{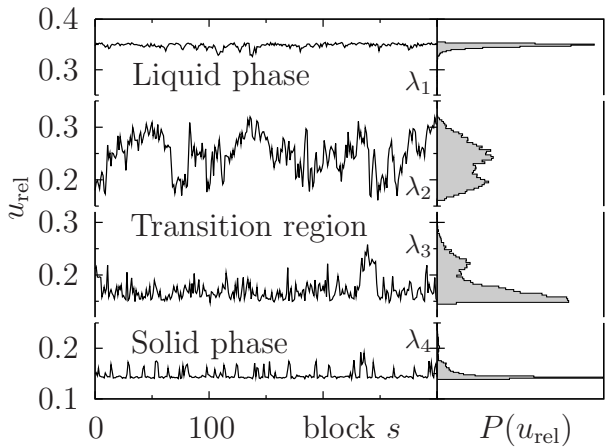}
 \caption{Left: Typical behavior of the block averaged IDF vs. block number $s$ for $N=8$ charged bosons in 2D for different coupling strengths $\lambda$: $\lambda_1=14$, $\lambda_2=22$, $\lambda_3=26$ and $\lambda_4=30$. Each point is an average over a block of length $M=1000$. Right: Histograms show the probability $P$ of different values \ur{} averaged over a total of $9000$ blocks. Results are from PIMC simulations with $200\ldots500$ beads \cite{numbook} of system \eqref{h}.}
 \label{fig:1}
\end{figure}

{\bfseries Solution of the convergence problem of \ur.}
We solve this problem by sub-dividing the time sequence in $K$ blocks of equal length $M$ ($L=K\cdot M$) and compute the \emph{block averaged IDF} $\ur(s)$ according to Eq.~\eqref{ur} for each block $s$~\cite{ berry93,noteexclude} and its mean $\urm=K^{-1}\sum^K_{s=1}\ur(s)$. To suppress the influence of jumps to \urm{} in the solid regime, $M$ must be chosen small enough to restrict jump-related contributions to a small number of blocks and, at the same time, large enough to allow for convergence of contributions related to local vibrations. This choice does not influence the convergence of \urm{} in the liquid regime which is dominated by frequent jumps on a time scale comparable to that of local vibrations and, hence, well below $M$. We demonstrate the behavior of \ur(s) for a quantum phase transition of $N=8$ bosons in 2D, cf. Fig.~\ref{fig:1}. In the solid regime there are rare spikes of \ur(s) corresponding to occasional blocks containing one jump leading to a sharply peaked probability distribution $P(\ur)$. In the transition region, however, each block may ``catch'' from zero to a few jumps, so the fluctuations of \ur(s) increase and $P(\ur)$ broadens. Finally, in the liquid regime, jumps occur with an almost constant rate in every block, so the fluctuations of \ur(s) are small [$P(\ur)$ has  again a single sharp peak], while the mean is shifted to a higher value above $0.3$, typical for a liquid. 

From this we conclude that, in the vicinity of the melting transition, the width of the distribution $P(\ur)$ reaches a maximum. This behavior is well captured by the second moment of $\ur(s)$, i.e.~the {\em variance of the block averaged interparticle distance fluctuations} (VIDF)
\begin{equation}
	\sigma_{\ur}	= \frac{1}{K}\sum_{s=1}^{K} \sqrt{\avr{\ur^2(s)}-\avr{\ur(s)}^2}.
	\label{su}
\end{equation}
This allows us to obtain a reasonable estimate of the melting temperature $T_{u}^{\text{crit}}$ from the peak of $\sigma_{\ur}(T)$~\cite{notemelt}.
Note that \urm{} is sensitive to the jump frequency $\nu_j$, in contrast to \ur{} of Eq.~\eqref{ur}. The sensitivity does depend on the block length $M$: larger $M$ cause an increase of \urm{} (as discussed before) and shift the maximum of $\sigma_{\ur}$ to lower temperatures, cf. Fig.~\ref{fig:0}.b)--c).

Therefore, to properly choose $M$ an independent quantity is needed which should not require block averaging and be invariant with respect to particle exchanges and pair distance jumps. A quantity fulfilling these requirements is the total energy $E$ and its autocorrelation function,
\begin{equation}
 \Ecorr(k)=\frac{\sum_{i=1}^{L-k} \left(E_{i+k}-\avr{E}\right)\left(E_{i}-\avr{E}\right)}{(L-k)\left(\avr{E^2}-\avr{E}^2\right)}.
 \label{e}
\end{equation}
We found that the decay rate of $\Ecorr(k)$ varies non-monotonically with temperature where the slowest decay is observed just in the transition region, cf. the example shown in Fig.~\ref{fig:0}.e). This suggests that the correlation time, $k_\text{corr}(T)=\sum_k \Ecorr(k,T)$, cf. Fig.~\ref{fig:0}.d), is sensitive to thermal melting, allowing us to identify the melting temperature $T_{E}^{\text{crit}}$ from the maximum of $k_\text{corr}$. Comparing the values $T_{E}^{\text{crit}}$ and $T_{u}^{\text{crit}}(M)$ provides a straightforward way to identify the proper block length $M$.
In all cases of thermal melting we investigated agreement is found for $M$ in the range of $1000\ldots10000$, where the common definition of a Monte-Carlo step is used~\cite{notePIMC}. 

We mention that in the case of quantum melting the situation is more complex. Nevertheless, we found that the same range of $M$ seems appropriate here as well, however, the analysis requires to use a combination of different quantities such as the pair distribution or bond angular symmetry parameters etc.

\begin{figure}
 \includegraphics[width=0.45\textwidth]{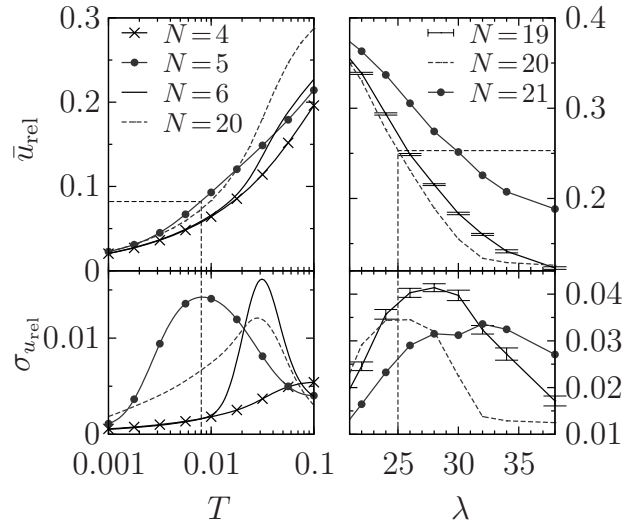}
 \caption{Mean value \urm{} (top) and second moment $\sigma_{\ur}$ (bottom) of the block averaged IDF for different particle numbers $N$. Left: temperature dependence of a classical 3D system (classical melting, classical MC simulations). Right: 2D quantum system, dependence on the quantum coupling parameter \relint{} (quantum melting, PIMC results). In both cases the block length equals $M=1000$. Dashed lines locate the critical values of $T$ (or $\lambda$) and $\ur^{\text{crit}}$.}
 \label{fig:2}
\end{figure}

{\bfseries Applications}. We have verified the behavior of the VIDF, $\sigma_{\ur}$, for a large variety of classical and quantum systems described by Eq.~\eqref{h} of various sizes and dimensionality. As a first illustration we show in Fig.~\ref{fig:2} (left side) MC results for a classical 3D system of $N=4...20$ particles the state of which is completely characterized by the temperature $T$. One clearly sees that in all cases \urm{} increases with $T$, but for small $N$ the reduction is very gradual, not allowing us to single out a ``melting temperature'' from \urm{}. At the same time, in all but one case $\sigma_{\ur}$ has a well pronounced peak at a certain $T$ which is identified as $T^{\text{crit}}$. Also, the critical value of the fluctations may be deduced from the peak position of $\sigma_{\ur}$ yielding $\ur^{\text{crit}}\approx 0.08\dots 0.16$ which is in good agreement with macroscopic classical Coulomb systems. Note the special case of $N=5$ showing a low value of $T^{\text{crit}}$ which is well known and explained by the low symmetry of this cluster~\cite{ludwig05}. While this behavior is hardly visible in \urm{} it is clearly detected by $\sigma_{\ur}$.

As a second example we consider quantum melting upon compression in a 2D system of spin polarized charged bosons at very low temperature close to the ground state. Calculations for particle numbers up to $N=60$ were done using PIMC simulations, for details see e.g.~\cite{numbook}. Right hand side of Fig.~\ref{fig:2} shows results for three cases, $N=19,20,21$, more examples are given in Ref. \cite{afilinov_bose07}. For large \relint, the particles are localized resembling a crystal as seen in Fig. \ref{fig:-1}. Decrease of \relint{} is associated with increasing wave function overlap and eventually quantum melting by tunneling of particles between lattice sites. Again we observe a gradual reduction of \urm{} when \relint{} is increased. In contrast, $\sigma_{\ur}$ has a pronouced peak which allows us to determine the critical value of \relint{} to $\relint\approx25\ldots30$ depending on the particle number. The corresponding critical fluctuations, $\ur^{\text{crit}}\approx 0.22\dots 0.25$, are again close to the value known from simulations of macroscopic Bose systems. 

These two examples are representative for the classical and quantum melting behavior of the system \eqref{h}, also for other pair potentials. All our calculations have confirmed the robustness and efficiency of the VIDF for the analysis of melting in small systems. We can now proceed and analyze the question what is the minimum system size to observe crystallization or melting? Our simulations have revealed that $\sigma_{\ur}$ has a maximum for particle numbers as small as $4$ in 2D and $5$ in 3D. In contrast, for $4$ particles in 3D, $\sigma_{\ur}$ shows a monotonic increase, see Fig.~\ref{fig:2} (top left). This is easily understood. The ground state of 4 (3) particles in 3D (2D) resembles an unilateral tetraeder (triangle) and has only a single interparticle distance. Thus, a jump (pair exchange) does not alter the distribution of pair distances, and $\sigma_{\ur}$ has no maximum.

In summary, we have proposed a novel quantity -- the {\em variance of the block averaged interparticle distance fluctuations} -- which is sensitive to fluctuations in finite systems. A maximum of $\sigma_{\ur}$ allows one to reliably detect the existence of structural changes which are analogous to solid-liquid phase transitions in macroscopic systems. It further directly yields a consistent estimate of the melting point \cite{notemelt} and the critical fluctuations $\ur^{\text{crit}}$ in classical and quantum systems, thereby curing the sensitivity and convergence problems of the conventional distance fluctuation parameters. 
While for classical systems the energy autocorrelation function $C_E$ allows for a calibration of the block length, this does not work for quantum melting where further 
analysis is required. Also, it remains an interesting question to analyze the behavior of $\sigma_{\ur}$ in other finite systems, including atomic clusters or homopolymers etc, as well as in time-dependend simulations (such as molecular dynamics). 
Finally, in the case of strongly inhomogeneous macroscopic systems where melting is known to proceed via a sequence of different processes, the VIDF should allow for a deeper insight and a space-resolved analysis of the fluctuations.

We thank Ch. Henning for helpful discussions.
This work is supported by the Deutsche Forschungsgemeinschaft via SFB-TR 24


\begin{thebibliography}{26}
\bibitem{notesurface}  Further, due to the strong radial increase of the confinement surface melting effects are strongly reduced.
\bibitem{noteurel} \ur{} accounts only for the distribution of pair distances and not the jump frequency.
\bibitem{noteexclude} Another solution is the exclusion of jump related contributions to \ur{}. However, in case of the system (\ref{h}), this would erase the ``phase'' information and is, therefore, not used.
\bibitem{notemelt} We do not imply that the system melts exactly at this temperature as ``melting'' is a gradual process in finite systems. The present approach allows for a systematic and consistent study of the entire crystal phase boundary.
\bibitem{notePIMC} A Monte-Carlo step is defined as $N$ displacement attempts in a $N$-particle system. In path-integral Monte-Carlo (PIMC) simulations used for quantum systems also different move types like bisectioning or end-point permutation are possible, cf.~\cite{binder, numbook}
\end{thebibliography}
\end{document}